\pdfoutput=1
\documentclass[11pt,twocolumn]{article}
\usepackage{graphicx}
\usepackage{epstopdf}
\usepackage{hyperref}

\usepackage[utf8]{inputenc}      
\usepackage[T1]{fontenc}         
\usepackage{lmodern}             
\usepackage{amsmath,amssymb,amsthm} 
\usepackage{graphicx}            
\usepackage{hyperref}            
\usepackage{geometry}            
\usepackage{authblk}             
\usepackage{cite}                
\usepackage{microtype}           

\title{Converting coherence into work with a fully quantum  engine}
\author[1,2]{MinSik Kwon}
\author[3]{Tobias Denzler}
\author[1]{Rouven Maier}
\author[1]{Vadim Vorobyov}
\author[1]{Durga Bhaktavatsala Rao Dasari}
\author[3]{Eric Lutz}
\author[1,2]{J\"org Wrachtrup}
\affil[1]{3rd Institute of Physics, ZAQuant, IQST, University of Stuttgart, Stuttgart, Germany}
\affil[2]{Max Planck Institute for Solid State Research, Stuttgart, Germany}
\affil[3]{Institute for Theoretical Physics I, University of Stuttgart, Stuttgart, Germany}
\begin{document}
\maketitle
\textbf{Heat engines convert thermal energy into mechanical work. We here report the experimental realization of a fully quantum engine that  converts quantum coherence into work. A single solid-state spin in diamond is fueled by a coherent bath and cyclically stores energy in a spin quantum battery. We establish quantum-enhanced performance  by showing that almost 200$\%$ more work is produced after a few cycle compared to the corresponding classical engine. We obtain concrete criteria for successful coherence-to-work conversion, and highlight the importance of a coherent motor-battery interaction. This device harnesses nonclassical features during all stages of its cycle, and demonstrates  the functionality of a nanomachine whose parts are all quantum coherent.}

Heat engines have been the centerpiece of classical thermodynamics since the very  beginning. Generating motion by cyclically converting  thermal energy into mechanical work, they have been instrumental in the investigation of energy conversion mechanisms \cite{cen01}. A heat engine is made of four distinct parts \cite{kos17}:  a working medium (or motor), two (cold and hot) heat reservoirs, and a  battery (or flywheel) that collects the produced work \cite{cen01}.  Such a work output device couples the engine to the outside world, and is therefore essential for its functionality.  Thermal machines have recently been  miniaturized to the  nanoscale \cite{ros16,lin19,kla19,ass19,pet19,hor20,bou20,kim22,ji22,koc23}, where  quantum properties, such as coherent superpositions of states \cite{str17}, are expected to lead to unusual thermodynamic features \cite{scu03,dil09,scu11,uzd15,fra20}. In particular, coherent reservoirs, combined with the conversion of quantum coherence into mechanical work,  have been predicted to  enhance overall engine performance \cite{scu03,dil09,scu11,uzd15,fra20,ergoDu}. This novel nonclassical conversion mechanism would  deposit more energy into the battery than a corresponding classical engine.  So far, only partial quantum heat engine operation has been reported \cite{kla19,ass19,pet19,bou20,hor20,kim22,ji22,koc23}.  {For instance, machines with a coherent working medium \cite{kla19,ass19,pet19,koc23},  a quantum reservoir \cite{bou20}, or a quantum system-bath coupling \cite{ji22}, but all without any flywheel, have been implemented. Furthermore, a classical motor coupled to a quantum load and classical baths has been described in Ref.~\cite{hor20}, while a quantum engine with a coherent bath but a classical piston has been realized in Ref.~\cite{kim22}}.

 {Realizing a fully quantum machine that preserves coherence and correlations between its working medium and  battery over multiple cycles is challenging, on any experimental platform, owing to the detrimental effect of decoherence \cite{zur03}. In particular, the cyclic reset of the state of the working medium to the  reservoir states after their interaction  should not erase the coherence of the battery. No generic solution to this general problem, which  also hinders the development of quantum networks, where the  readout and reset of one memory qubit  should not decohere the other qubits \cite{rei16,bra22}, is currently available.} On the other hand, the crucial necessary requirements for successful cyclic coherence-to-work conversion have not been experimentally explored yet. 

 We here present the experimental realization of a fully quantum, single spin-$1/2$  engine powered by nonclassical reservoirs. We concretely implement a quantum Otto-like cycle, a quantum generalization of the common four-stroke car engine \cite{kos17}, using the electron spin of a single nitrogen-vacancy (NV) center in diamond \cite{doh13} as the working substance and a strongly coupled carbon nuclear spin as the battery. Both spins can be controlled to a high degree using microwave and radiofrequency fields  \cite{doh13}. Cold and hot reservoirs, with tunable amount of coherence in the energy basis, are created by employing additional nuclear spins and microwave fields. We moreover ensure coherence-preserving reset by carefully engineering the interaction between the electron and carbon spins by operating close to the magic angle where the $zz$-component of their interaction vanishes \cite{cal91}. We observe coherence-to-energy conversion along with quantum work storage  over multiple coherent  cycles.  We additionally obtain criteria for successful coherence-to-work conversion and underscore the importance of a coherent motor-battery interaction. Finally, we demonstrate quantum-enhanced performance with larger work output than classically possible during coherent multi-cycle engine operation, and establish the quantumness of motor and battery, as well as their joint state by measuring single and two-spin correlators \cite{how06,hanson1}, and the concurrence \cite{woo98}.

\begin{figure*}[t]
\includegraphics[width=0.98\textwidth]{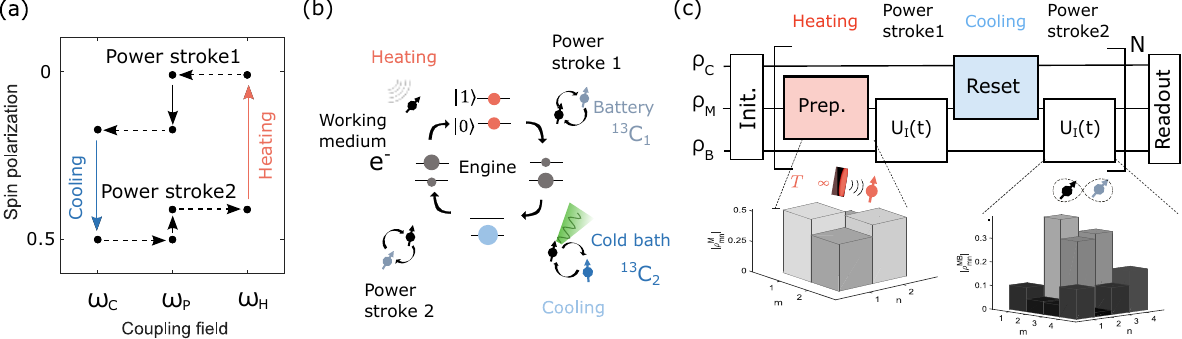}
\caption{\textbf{Coherent engine}. a) Otto-like cycle for a spin-1/2 quantum working medium (motor) consisting of four steps: coherent heating, expansion, cooling and compression. The expansion and compression phases of the engine represent the  {adiabatic interaction stages}, wherein the motor coherently exchanges energy with a  spin-$1/2$ quantum battery, resulting in a power stroke. b) The quantum engine is experimentally implemented using the electron spin of a nitrogen-vacancy (NV) center in diamond (motor) and two nuclear carbon atoms ($\textsuperscript{13}$C$_{1}$ for the battery and $\textsuperscript{13}$C$_{2}$ for the cold bath). Heating is achieved by applying  microwave fields, whereas coherent flip-flop interactions between motor and battery is realized with the help of Hamiltonian engineering. c) The subpanels display the measured one and two-qubit density matrices during initialization and power stroke stages that verify the nonclassicality of the medium state  as well as the joint medium-battery state (see SI). }
\centering
\label{f1}
\end{figure*}
 
\textit{Fully coherent engine.} We begin by  analyzing a minimal model of a coherent  engine in order to determine specific criteria for  coherence-to-work conversion and quantum-enhanced performance. {To that end, we consider an Otto-like cycle for a  spin-1/2 working medium with Hamiltonian $H_M=(\hbar\omega_M/2) \sigma_M^z$, where $\omega$ is  the frequency and  $\sigma^{x,y,z}$ are  the usual  Pauli operators.} The cycle  consists of four   steps (Fig.~\ref{f1}a): 1) Coherent heating: a coherent bath at infinite temperature prepares the  qubit in  the state $\rho_{M} = \frac{1}{2}{I}+ P_M^x \sigma^x_M$, where the polarization $P_M^x$ sets the strength of the  coherence along the $x$-axis and ${I}$ is the identity operator. The state  $\rho_{M}$ describes a classical fully mixed two-level system for $P_M^x =0$; 2)  {Adiabatic interaction (expansion)}: the working medium coherently interacts with a qubit battery via an exchange (flip-flop) interaction of the form  $H_I =  g(\sigma_M^+\sigma_B^- + \sigma_M^-\sigma_B^+$), with amplitude $g$ \cite{pulsepol}, where $\sigma^\pm = \sigma^x \pm i\sigma^y$. During this  first power stroke, work is done on the battery; 3) Cooling: the working medium is equilibrated to a cold bath at zero temperature; 4)  {Adiabatic interaction (compression)}: another coherent flip-flop interaction with the battery implements a second power stroke.

 \begin{figure*}[th]
\includegraphics[width=1.0\textwidth]{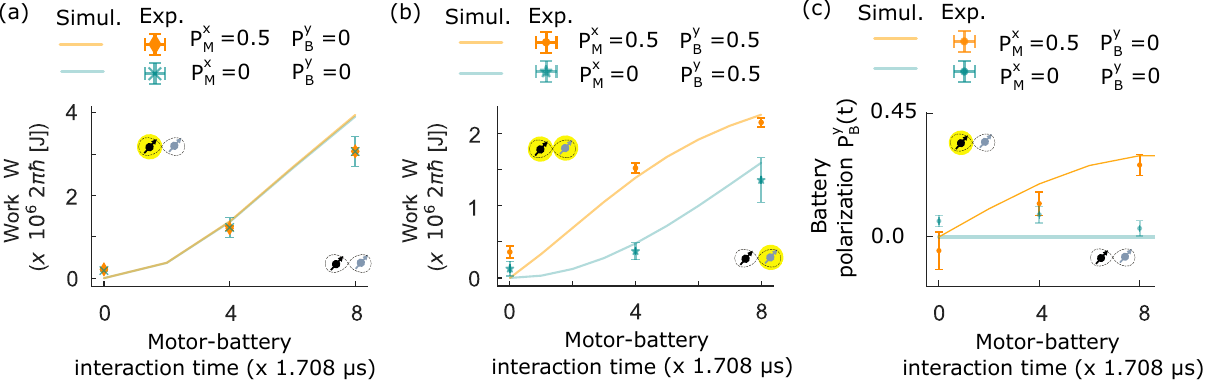}
\caption{\textbf{Engine performance over one cycle.} a) The work $W$ deposited into the battery during the first cycle, with increasing duration of the power stroke. Work is the same for incoherent (blue) and coherent (orange) heating, indicating that coherence-to-work conversion is ineffective when the battery is classical. b) By contrast, coherence is converted into work when the battery is also quantum, demonstrating quantum-enhanced work production (coherent spins are highlighted in yellow). c) The battery spin polarization $P_B^y(t)$ increases with the interaction time, showing that the quantum motor-battery coupling transfers coherence from the working medium  to the flywheel.  Good agreement with theoretical simulations (solid lines) is observed in all cases. Error bars correspond to one standard deviation.}
\centering
\label{f3}
\end{figure*}
Insight into the engine dynamics may be gained by evaluating the  energy deposited into the battery during the first cycle by analytically  computing the work performed during the   two power strokes. We write the initial density operator of the battery as $\rho_B = \frac{1}{2}I + \vec{P}_B \cdot \vec{\sigma}_B$, where $\vec{P}_B$ is the spin-polarization vector  on the Bloch sphere and $\vec{\sigma}_B$ the corresponding Pauli vector \cite{ben06}. Energy exchange between motor and battery is determined by $\Delta P_B^z$, while coherence transfer is characterized by  $\Delta P_B^{x,y}$. Taking an initial separable joint state $\rho_M\otimes\rho_B$ and exploiting the unitary evolution $U_I(t)$ generated by the interaction $H_I$, we obtain the work stored in the battery after one full cycle,  {defined as the mean energy change  of the battery} (Supplementary Information)
\begin{equation}
\label{1}
W =  \hbar\omega_B (P^x_M P^y_B \sin\theta  \cos^3\theta  + P^z_B(\cos^4\theta-1)+\sin^2\theta),
\end{equation} 
with the angle $\theta = gt$ and the interaction time $t$ during each power stroke. Expression \eqref{1} contains a purely classical term, $W_{c}= \hbar\omega_B (P^z_B(\cos^4\theta-1)+\sin^2\theta)$,  and a quantum contribution, $W_{q}=  \hbar\omega_B P_M^x P^y_B \sin\theta  \cos^3\theta$, that depends on the  coherence parameters $P_M^x$ and $P^y_B$ of medium and battery. The battery frequency $\omega_B$ has contributions both from the external B-field and the hyperfine interaction (see Supplementary Information). As expected, $W_q$ vanishes for $P_M^x=0$, that is, for an incoherent  thermal hot bath. Surprisingly, it also vanishes for a classical battery $P^y_B=0$, indicating that quantum-enhanced work output does not occur in the absence of a quantum battery. This leads us to the formulation of a first criterion for successful coherence-to-work conversion: both working medium and flywheel should exhibit  coherence in their energy basis.

\textit{Experimental realization.} We implement the above minimal model of a coherent engine by using the central electron spin of a NV center in diamond as working medium (Fig.~\ref{f1}b).  NV center systems offer excellent control of their states and exhibit very long spin coherence times \cite{doh13}.  The hot bath is emulated with the photon field (green laser) used to
initialize and read out the NV center electron spin (Supplementary Information).  {For the implementation of a single cycle operation of the engine (Fig.~\ref{f3}), the cold bath and the battery  are chosen as two  carbon $^{13}$C nuclear spins that are coupled to the central electron spin with respective (longitudinal) strengths   414kHz and 90kHz. The diagonal states of hot and cold baths are respectively given by $\rho_ H= \text{diag}(0.485,0.515)$  and $\rho_ C= \text{diag}(0.03,0.97)$.}   {We next realize the cooling step   using an effective SWAP operation between the cold reservoir (in its ground state) and the working medium.} The coherence $P_M^x$ is added through microwave fields that allow for the rotation of the spin. The energy cost for this rotation in comparison to the final work output is much smaller and hence neglected (Supplementary Information). We further  implement the coherent flip-flop interactions with the help of Hamiltonian engineering \cite{pulsepol} (Fig.~\ref{f1}c). The performance of the machine is evaluated by measuring the $z$-polarization $P^z_B$ of the battery, giving direct access to the energy stored in the battery; the coherence $P^y_B$ is determined by rotating the Bloch vector along the $z$-axis (Supplementary Information). The long coherence time of the nuclear spins allows for the storage of populations and coherences during multiple repetitions of the engine cycle.

We first characterize the quantumness of the working medium state $\rho_M$   after  coherent heating by measuring the spin expectation values $\langle \sigma_M^j\rangle$, $j=(x,y,z)$ \cite{how06}. Figure~1c) clearly shows  quantum coherence along the $x$-direction. Likewise, we identify the nonclassicality of the joint medium-battery state $\rho_{MB}$ after the flip-flop interaction $H_I$ by evaluating  the two-qubit correlations $\langle \sigma_M^j\sigma_B^j\rangle$ \cite{hanson1}. Any nonzero value seen in Fig.~1c) confirms the presence of off-diagonal elements in the two-qubit density matrix (Supplementary Information).  {We  additionally determine a concurrence of 0.4 (Supplementary Information), which shows that the two qubits are entangled \cite{woo98}}.

Figure~\ref{f3} examines the transfer of energy and coherence between motor and battery during the first cycle as a function of the interaction time. For an initial  classical battery ($P_B^y=0$), the amount of work $W$ produced by the engine increases as a function of the interaction time; it is  the same whether the working medium contains  coherence (orange) or not (blue) (Fig.~\ref{f3}a). In this case, coherence-to-work conversion is not effective, as predicted by the analysis of the theoretical model \eqref{1}. By contrast, for an initial quantum battery  ($P_B^y=0.5$),  {prepared  by a RF pulse}, significantly more work is deposited into the battery, when the working medium is thermalized by  a coherent bath, demonstrating successful coherence-to-work conversion (Fig.~\ref{f3}b). Figure~\ref{f3}c) additionally shows that the coherent exchange interaction $H_I$ is able, due to his entangling nature, to also transfer coherence from the motor to the battery, even when the battery is initially classical, as seen by the steady increase of the polarization  $P_B^y(t)$ with time; on the other hand,  $P_B^y(t)$ remains unchanged for a classical working medium. We shall exploit this important property in the following. In all instances, we have good agreement between data (symbols) and numerical simulations (solid lines) (Supplementary Information).  {We additionally stress that the work enhancement observed in Fig.~2 is not a simple coherent transfer of polarization in the energy basis as known earlier \cite{pulsepol}, but is a careful synchronization of the motor-battery interaction $H_{I}$ with the precession of the target spin along the $I^x_B - I^y_B$ plane caused by the initial coherence of the battery spin along the $y$-axis. Since the flip-flop interaction ($H_{I}$) is phase sensitive, the energy (polarization) transfer can be controlled. Another advantage of the engineered flip-flop interaction is its reliance on a rapid, MW-only approach, which dramatically reduces control times by three orders of magnitude relative to those requiring RF-driven control, preserving quantum coherence far easier throughout the engine cycles (Supplementary Information).}

\begin{figure*}[th]
 \includegraphics[width=.9\textwidth]{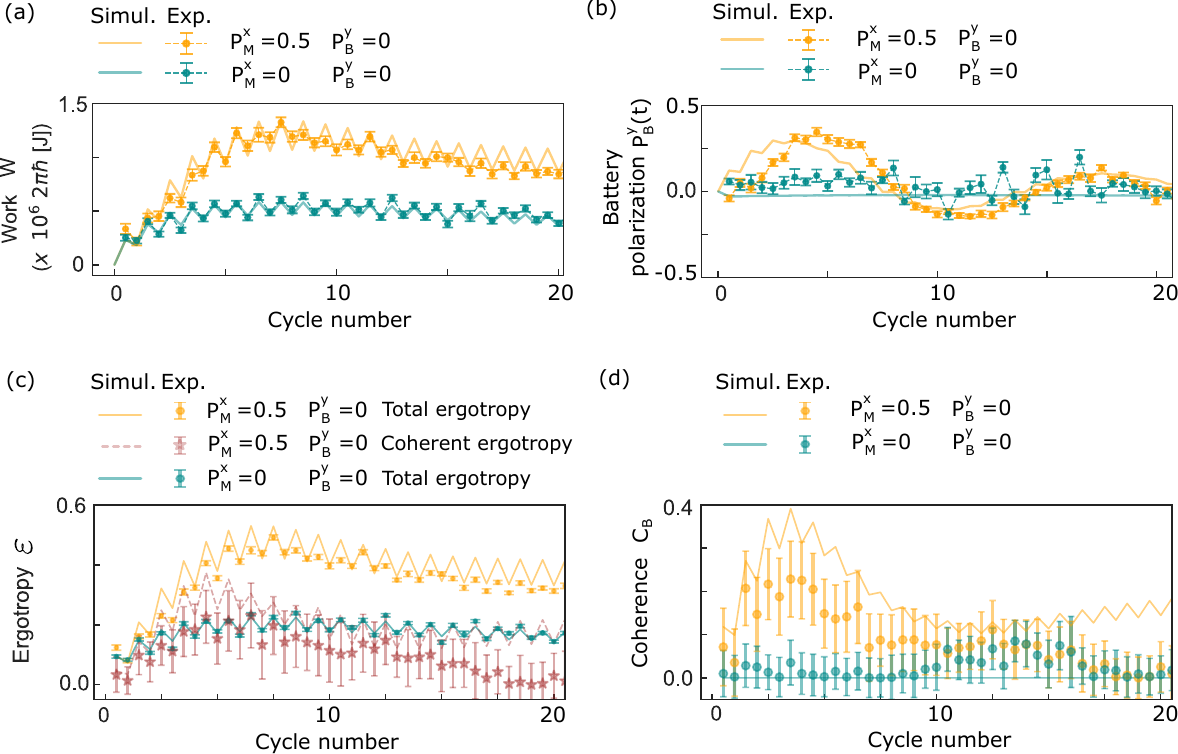}
\caption{\textbf{ Performance of the coherent engine over many cycles}. a) After the first few cycles,  work production is significantly enhanced for coherent heating (orange) compared to incoherent heating (blue), demonstrating coherence-to-work conversion and quantum advantage, as well as effective work storage. Maximum work production is achieved after $N=8$ cycles, with an enhancement of almost 200$\%$ compared to the classical case,   {through coherence-preserving reset of the hot and cold baths}. b) Coherence-to-work conversion is facilitated by the transfer of coherence from the  motor to the battery, leading to an increase of the battery polarization $P_B^y$. c) The ergotropy $\cal E$ of the battery is equally increased in the presence of a coherent hot bath. The coherent part of the ergotropy ${\cal E}_q$ is shown in purple. d) The coherent ergotropy is related to the relative entropy of coherence $C_B$ of the battery, which quantifies its amount of quantum coherence, and exhibits a similar behavior as a function of the cycle number. Error bars correspond to one standard deviation.}
\centering
\label{f4}
\end{figure*}
The multi-cycle performance of the engine is displayed in Fig.~\ref{f4}.  {In order to protect the coherence of the battery from  the reset of the working medium during during a multicycle operation  we carefully chosen the battery as a weakly coupled $^{13}$C spin that has vanishing longitudinal coupling  $(A_{zz} \sim 0)$ and a very low transverse coupling ($A_{zx} \sim 17$kHz) \cite{vor23}, thus achieving state-of-the-art reset robustness \cite{rei16,bra22} (Supplementary Information). Furthermore, the direct optical reset via the green laser is considered as the cold bath action in the engine cycle}. As above, we first take the battery to be initially in a classical state ($P_B^y=0$). In line with the previous discussion, the produced work $W$ is mostly the same for coherent and incoherent working media for the first few cycles. Remarkably, we observe that there is a sharp quantum enhancement  of the work output (orange), of almost 200$\%$ after 8 cycles (Fig.~\ref{f4}a). In this instance, $2/3$ of the work is gained from quantum coherence, while only $1/3$ stems from heat. This quantum advantage is facilitated  by the  initial transfer of coherence from motor to battery (Fig.~\ref{f4}b), which makes  the latter nonclassical $P_B^y(t)\neq 0$; this enables, in turn, the conversion of quantum coherence into work and increased work production. This leads us to a second condition for successful coherence-to-work conversion: in the presence of a classical battery, the motor-battery interaction should be coherent, and permit coherence transfer in the energy basis between the two. The  engine exhibits a robust quantum advantage (up to 20 cycles) after which experimental imperfections (dominated by the initialization of the cold bath) and  environmental decoherence degrade its performance (Supplementary Information). We emphasize that since the cold bath is at zero temperature and the hot bath at infinite temperature, the work produced by the incoherent  engine is the largest possible for a classical two-level Otto cycle with all other parameters fixed.

Figure~\ref{f4}c) further presents the ergotropy $\cal E$ of the battery,  which corresponds to the maximum amount of work which can be extracted from it via a cyclic unitary \cite{all04,niu24}. Ergotropy is a key quantity in quantum thermodynamics that only vanishes for passive states like thermal states; it usually differs from a system's mean energy. It is given by ${\cal E} = \text{Tr}[(\rho_B^N - \tilde\rho_B^N)H_B]$, where $H_B = \left({\hbar\omega_B}/{2}\right)\sigma^z_B$, and $\rho_B^N$ its   density operator after cycle $N$, and $\tilde \rho_B^N$ the corresponding passive state with population probabilities ordered in increasing order \cite{all04,niu24}. The ergotropy $\cal E$ exhibits a behavior similar to that of the work $W$ deposited into the battery, showing a strong quantum boost in the energy that can be gained from it, when the number of cycles increases. This boost can be quantified more precisely with the help of the  coherent part ${\cal E}_q$ of the ergotropy (purple), defined as the part of the  extractable work which cannot be obtained by means of incoherent operations \cite{fra20}. The coherent entropy is intimately related to the  presence of quantum coherence in the state of the battery, as expressed by the relative entropy of coherence, $C_{B} (\rho) = S (\rho^d_B) - S (\rho_B)$,  where where $S(\rho_B) = - \text{Tr} \rho_B \ln \rho_B$ is the von Neumann entropy and $\rho_B^d$ is the diagonal  operator in the  energy basis \cite{str17}. The relative entropy of coherence grows with the number of cycles (Fig.~3d), before it is degraded due to experimental errors and decoherence, like the polarization $P_B^y(t)$.

\section*{Conclusion}\label{sec13}
The   19th century discovery that heat is a form of energy and that it can be converted into work is a cornerstone of classical thermodynamics. Likewise, the conversion of coherence into work may be regarded as a pillar of modern quantum thermodynamics. We have successfully demonstrated such coherence-to-work conversion, as well as  effective work storage,  using a cyclic  engine whose parts are all quantum. The engine does not solely convert heat into work, as classical heat engines do, but also (and mostly) the quantum coherence provided by a coherent hot reservoir.   {To that end, we have implemented a working medium-battery interaction that allows fast cyclic readout and reset of the working medium that preserves the quantumness of the battery. Such versatile, and scalable, step is not only essential for the realization of fully coherent cyclic machines, but may also be useful  for other quantum information processing applications, including quantum networks \cite{rei16,bra22}.} We have furthermore characterized the corresponding quantum-enhanced power output using the  ergotropy, and highlighted the importance of a coherent motor-battery interaction. 
The choice of the motor-battery coupling thus plays  a vital role in determining whether quantum coherence can be effectively used or not. It also underlines the fact that the properties of quantum machines cannot be fully analyzed in the idle mode where they are not coupled to a battery, as was commonly done so far.  {We expect these findings to positively impact the development of quantum energy storage devices, such as quantum batteries \cite{cam24}, and of nanoscale energy converters with increased performance \cite{lin13}.} Augmenting the number of spins in every component of the device and exploiting the full many-body dynamics of the system should allow us to further harness nonclassical features, such as coherence and correlations, to design quantum enhanced machines.

\section*{Acknowledgements}
We acknowledge  support from DFG (Project No:FOR2724), the German Ministry of Education and Research for the project InQuRe (BMBF, Grant agreement No~16KIS1639K), QR.X (BMBF, Grant agreement No~16KISQ013),  from the European Commission for the Quantum Technology Flagship project QIA (Grant agreements Nos~101080128 and 101102140), the Baden-Wuerttemberg Stiftung for the project SPOC (Grant agreement No~GRK2642). We would also like to acknowledge the fruitful discussions and comments from K. Micadei, J. Meinel, C. Ho and T. Jaeger.




\end{document}